\documentclass{elsart5p}

\usepackage{graphics}
\usepackage{graphicx}
\usepackage{amssymb}

\begin{document}

\begin{frontmatter}



\title{Novel Photo-Detectors and Photo-Detector Systems}


\author{M. Danilov}

\address{Institute of Theoretical and Experimental Physics,\\ B.Cheremushkinskaya 25, 117218 Moscow, Russia\\ e-mail danilov@itep.ru}

\begin{abstract}

Recent developments in photo-detectors and photo-detector systems are
reviewed. The main emphasis is made on Silicon Photo-Multipliers
(SiPM) - novel and very attractive photo-detectors. Their main
features are described. Properties of detectors manufactured by
different producers are compared. Different applications are discussed
including calorimeters, muon detection, tracking, Cherenkov light
detection, and time of flight measurements.
% A comparison with other photo-detectors is made.

\end{abstract}


\end{frontmatter}


\section{From PMT to SiPM}
\label{sec1}

Vacuum Photo-Multipliers (PMTs) are the most popular
photo-detectors. They have high sensitivity, single photo-electron
resolution, high counting rate, large area, good time
resolution. There is an enormous experience in PMT applications in
different fields. However PMTs have also drawbacks: sensitivity to
magnetic field, large size and low granularity, low quantum efficiency
($QE$), need of high voltage. They are also quite expensive. These
drawbacks can be partially cured. Multi-anode PMTs (MAPMT) offer
higher granularity. Micro Channel Plate PMT can work to some extent in
magnetic fields. Recently PMTs with high ($50\%$) quantum
efficiency have been developed \cite{PMT}.

However, only solid state detectors can provide a cardinal solution of
the problems. They are insensitive to magnetic field and compact. They
have very high quantum efficiency and granularity. Solid state
photo-detectors can be much cheaper than PMTs.

The simplest photo-detector is a PIN photo-diode. It has no
amplification and therefore it is very stable. PIN photo-diodes have a
high ($\sim80\%$) $QE$ well matched to the CsI(Tl) emission spectrum
which peaks at $\lambda\sim550\,$nm. Therefore they have been used in
large quantity in many calorimeters including CLEO, BELLE, BaBar, and
GLAST. However a thick ($\sim300\,\mu$m) sensitive layer leads to a
large Nuclear Counting Effect. Charged particles crossing the
sensitive layer produce a large number 
%($\sim30000$) 
of electron-hole
pairs and mimic a large energy deposition in a scintillator. Absence
of amplification prevents usage of PIN photo-diodes with low light
yield scintillators.

These two problems are solved in Avalanche Photo Diodes (APD). In APD
photo-electrons (p.e.) are produced in a thin ($\sim6\,\mu$m) sensitive
layer amplified in avalanches at a p-n junction.  About 120 thousand
APDs are used in the CMS calorimeter\cite{CMS}. Excellent resolution
has been achieved in spite of a low photon yield of PbWO4
crystals. Because of avalanche amplification APDs have a large Excess
Noise Factor (ENF) which grows with the amplification. Voltage and
temperature sensitivities of the amplification also grow with the
amplification. Therefore it is difficult to operate APDs at
amplifications above a few hundred.

At a high over-voltage ($\Delta V$) the avalanche amplification transforms into a
Geiger discharge. In this mode a photo-diode response does not depend
on a number of initial photo-electrons. However it is possible to
restore the proportionality of the response to the initial number of
p.e. by splitting a photo-diode into a large number of independent
pixels connected to the same output. The number of fired pixels is
proportional to the number of initial p.e. as long as it is small in
comparison with the total number of pixels in a photo-diode. For larger
signals the response becomes nonlinear and saturates at the total
number of pixels in the photo-diode.
% So the dynamic range is
%determined by the total number of pixels.  
Such multi-pixel
photo-diodes working in the Geiger mode have been developed 
in Russia\cite{SiPM}. Now they are produced by
many companies which use different names for their products: SiPM, MRS
APD, MPPC, MAPD, etc. We will use a generic name SiPM for all of them.

\section{ SiPM properties}
\label{sec2} 

We will discuss SiPM properties using as an example the MEPhI-Pulsar (MEPhI)
SiPM. There is by far the largest experience in using such SiPMs in
real experiments. About 8 thousand of them were used in the CALICE hadron
calorimeter prototype for ILC\cite{AHCAL,Felix} which was tested during the
last 3 years at CERN and FNAL.

MEPhI SiPM is a matrix of 1156 pixels
with the size $32\times32\,\mu$m$^2$. 
%(see Fig.~\ref{SiPM})
%
%\begin{figure}[htbp]
%\includegraphics[width=9cm]{SiPM.eps}
%\caption{Photographs of the MEPhI SiPM.}
%\label{SiPM}
%\end{figure}
The SiPM sensitive area is $1.1\times1.1\,$mm$^2$. The pixels have
individual polysilicon quenching resistors of a few $M\Omega$
necessary to break off the Geiger discharge. SiPMs are reversely
biased with a voltage of about 50V which is about 3V higher than the
breakdown voltage ($V_{BD}$).

\subsection{Gain and Photon Detection Efficiency}

SiPM gain (G) is determined by a charge released in one pixel
discharge which is proportional to a pixel capacitance (C) and
$\Delta V$: $Q=\Delta V\times C$. Typical values of
$\Delta V\sim3\,V$ and $C\sim50\,fF$ lead to $Q\sim150\,$fC. So one
p.e. produces a signal of about $10^6$ electrons. This is very
similar to a usual PMT. The relative gain variation $\Delta G/G$ is
proportional to the relative $\Delta V$ variation. Therefore SiPMs
operated at smaller $\Delta V$s like Hamamatsu MPPCs have larger
gain sensitivity to the voltage variation and require a better voltage
stabilization. A decrease of temperature by $2^\circ C$ leads to the
decrease of the breakdown voltage of the MEPhI SiPM by $\sim0.1\,$V and
hence to the increase of the gain. Therefore it is desirable to keep
temperature variations small.

Photon Detection Efficiency ($PDE$) is a product of $QE$, a geometrical
efficiency ($\epsilon$), and a probability for a charge carrier to
initiate the Geiger discharge ($P_G$): 
$PDE = QE\times\epsilon\times P_G.$ 
$QE$ is about 80\% at $\lambda = 500\,$nm. The geometrical efficiency is
a fraction of a SiPM area which is sensitive. It decreases with the
decrease of a pixel size since the area of separating boarders between
pixels grows. The geometrical efficiency of modern SiPMs can be as
large as 70\% for $50\,\mu$m pixels. The probability of the Geiger
discharge increases with the $\Delta V$. It grows almost linearly at
small $\Delta V$s but then saturates. This leads to a similar behavior
of $PDE$.

\begin{figure}[htbp]
\includegraphics[width=8cm]{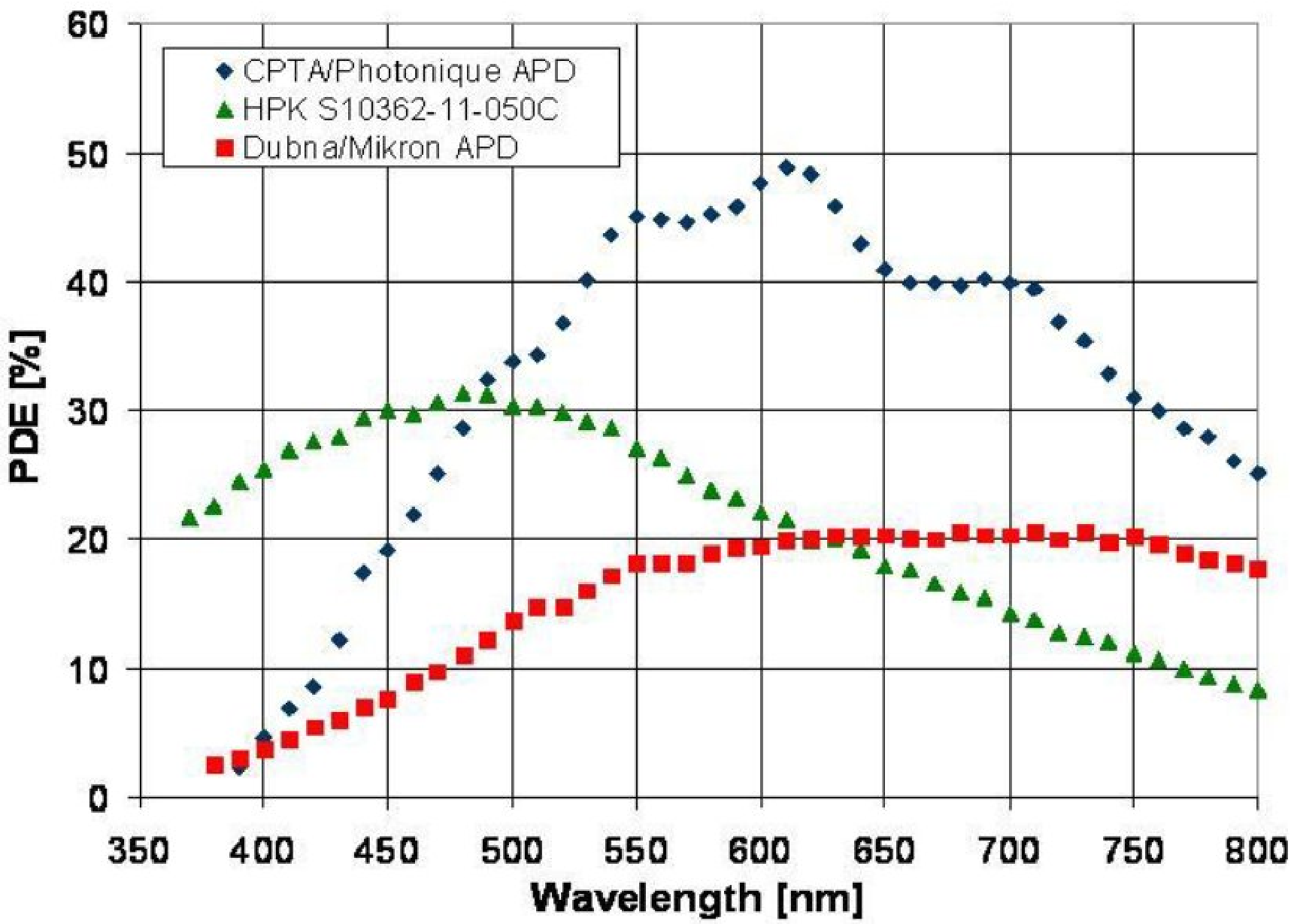}
\caption{A photon detection efficiency for different SiPMs.}
\label{musienko}
\end{figure}

Electrons have a much higher probability to trigger the Geiger
discharge in the p-n junction than holes. Therefore SiPMs with a n-p
structure are more sensitive to a green light than to a blue light. A
blue light (green light) is absorbed in the n-layer (p-layer); the
carriers which move to the p-n junction are holes (electrons) and the
probability to trigger the discharge is low (high).
In order to increase the sensitivity for blue light the n-layer should
be made as thin as possible. Another possibility is to use a p-n
structure. Fig.~\ref{musienko} shows examples of $PDE$ spectral
dependence for different SiPMs\cite{PDE}.
\subsection{After-pulses and cross-talk}

\begin{figure}[htbp]
\includegraphics[width=8cm]{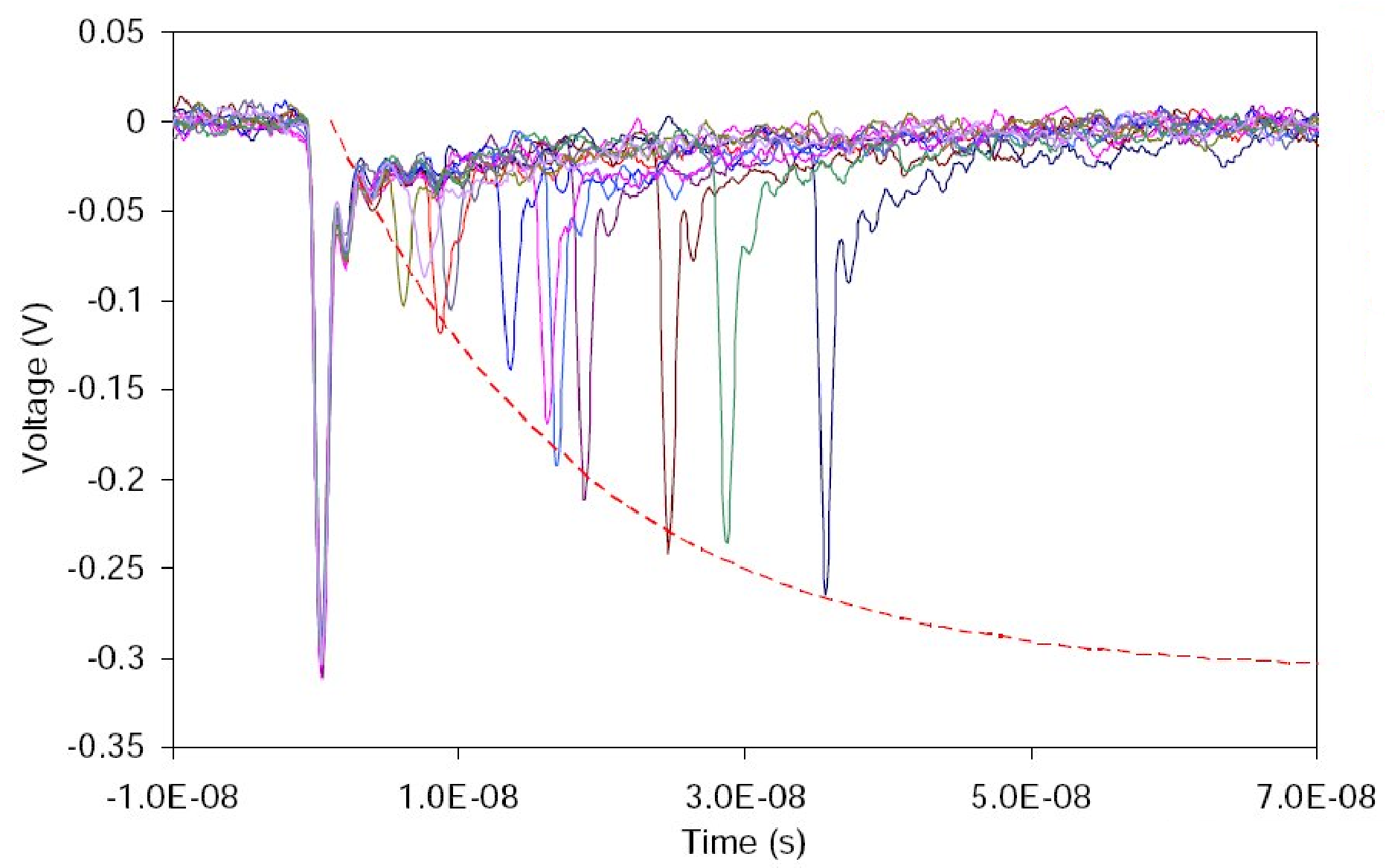}
\caption{Examples of after-pulses with different delays.}
\label{afterpulse}
\end{figure}

Some electrons and holes produced in the discharge can be trapped and
then released when the discharge is already quenched. This leads to
after-pulses if the pixel has sufficient time to recharge (see 
Fig.~\ref{afterpulse}\cite{piemonte}).
The after-pulses which come soon after the initial signal have smaller
amplitudes (see Fig.\ref{afterpulse}) since the pixel voltage is not completely
restored. Majority of after-pulses come soon after the initial signal
with a decay time of about 18ns\cite{T2K1}. However there is also a
fraction with a longer decay time $\sim90\,$ns. The decay times become
shorter at higher temperatures\cite{T2K1}. The after-pulse probability
is proportional to the number of electrons in the discharge (i.e. to
the gain) and the probability to trigger the Geiger
discharge. Therefore it grows roughly as the second power of the
$\Delta V$.

The pixel recovery time depends on the pixel capacitance and quenching
resistor ($R$). It can not be much smaller than $100\,$ns since the
quenching resistor can not be small.  For such small recovery times
the pixel can be fired more than once during a long light pulse coming
for example from a scintillator. This increases the dynamic range of a
SiPM. However it also makes the response dependent on the shape of the
light pulse.  This complicates the calibration procedures. Majority of
SiPM types have longer recovery time
%s from several hundred ns to
$\sim1\,\mu$s. The SiPM dead time is much smaller since the number
of pixels is large. Therefore SiPMs can tolerate high counting
rates. An LED signal was well seen in a scintillator strip with a SiPM
readout when we irradiated it with a $^{90}Sr$ source up to a counting
rate of $600\,$kHz ($I=3\mu$A).

Photons are created in the Geiger discharge with a rate of
$\sim3\times10^{-5}$/electron at $\lambda<1.1\mu$m\cite{phot/e}. Low
energy photons have a long absorption length up to $\sim 1\,$mm at
$\lambda = 1.1\,\mu$m and can produce p.e. in neighboring pixels.  This
leads to the inter-pixel cross-talk.
%Photons with $\lambda<
%700\,$nm have the absorption length below $5\,\mu$ and practically do
%not contribute to the cross-talk.  
Photo-electrons can be produced in the pixel active region or in the
bulk.  Therefore there is a prompt and delayed component in the
cross-talk\cite{mirzoyan}.  The prompt component can be suppressed by
trenches between pixels. The delayed component can be suppressed by an
additional p-n junction. This is nicely demonstrated by the MEPhI-MPI
(Munich) group. Such double cross-talk suppression allowed them to
reduce the cross talk to below 1\% level at the amplification of
$\sim2\times10^7$\cite{mirzoyan}.
%One should mention however that
%the studied SiPM has a large gap between pixels which reduces the
%cross-talk.
Modern industrially produced SiPMs have cross-talk values at working
$\Delta V$s of 5-10\%. The cross talk and after-pulses lead to the
increase of ENF which is however still much closer to 1 than in APD or
even PMT.

\subsection{SiPM response}

The SiPM response is a product of several factors. For small light
pulses it is given by

\begin{figure}[htbp]
\includegraphics[width=4.35cm]{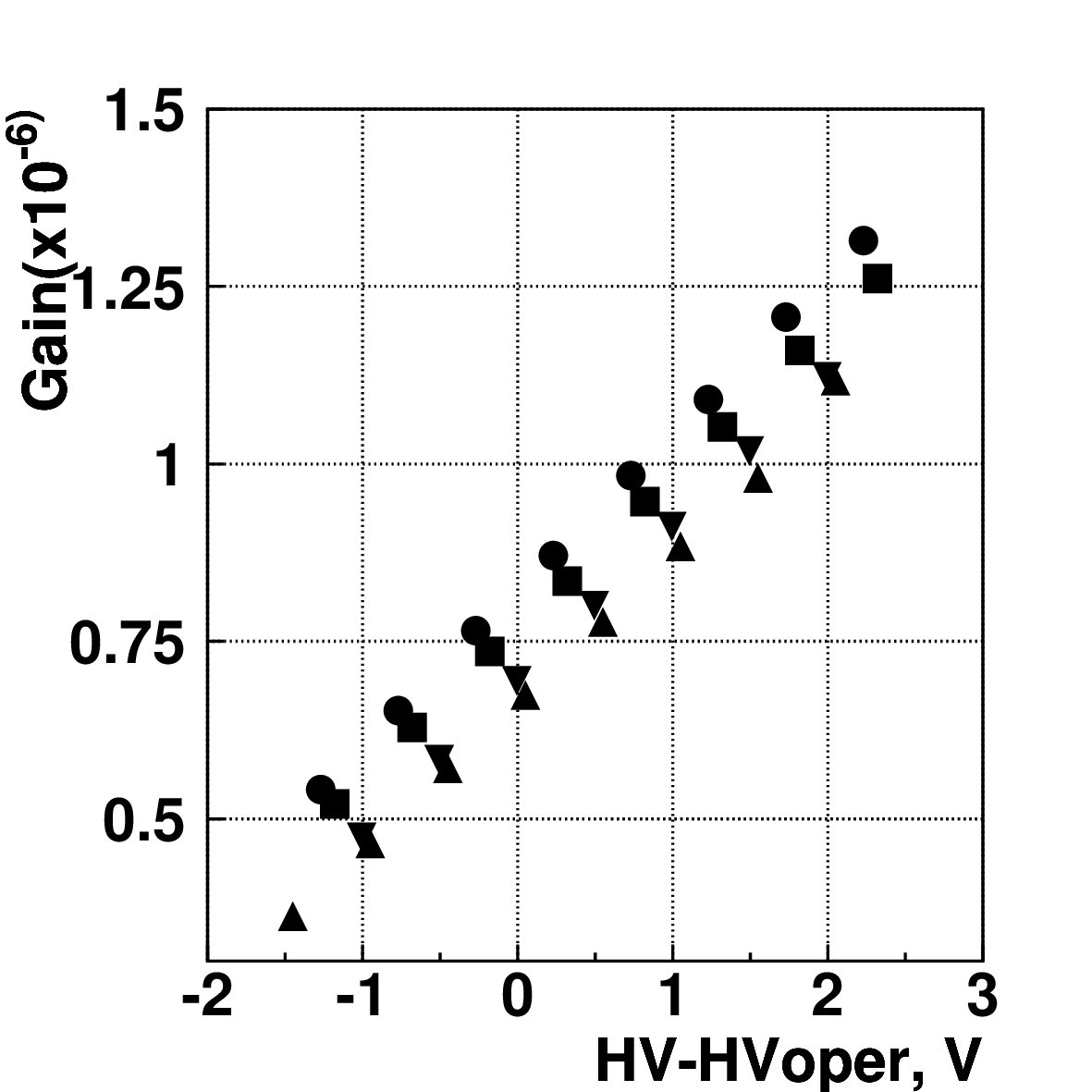}
\includegraphics[width=4.35cm]{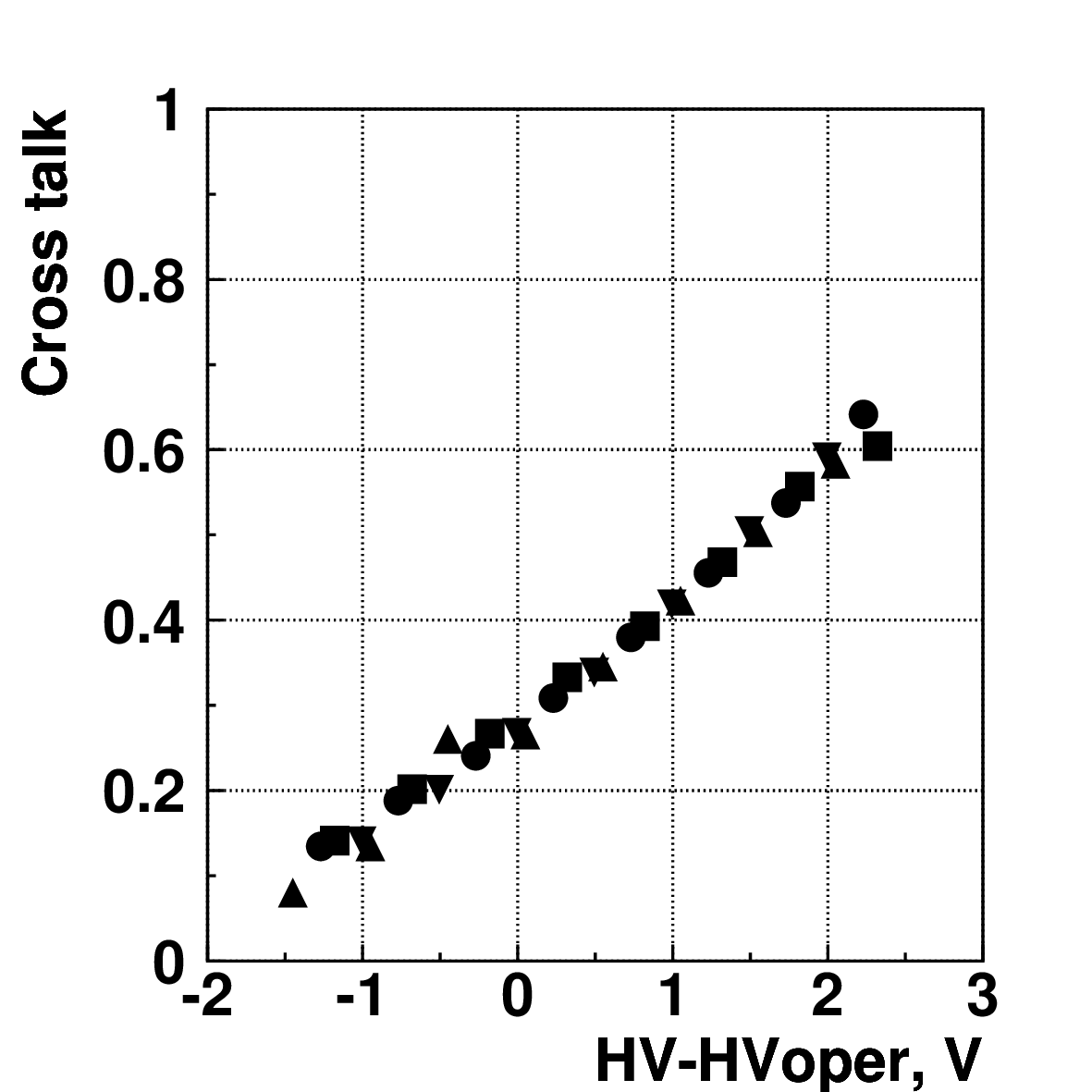}
\includegraphics[width=4.35cm]{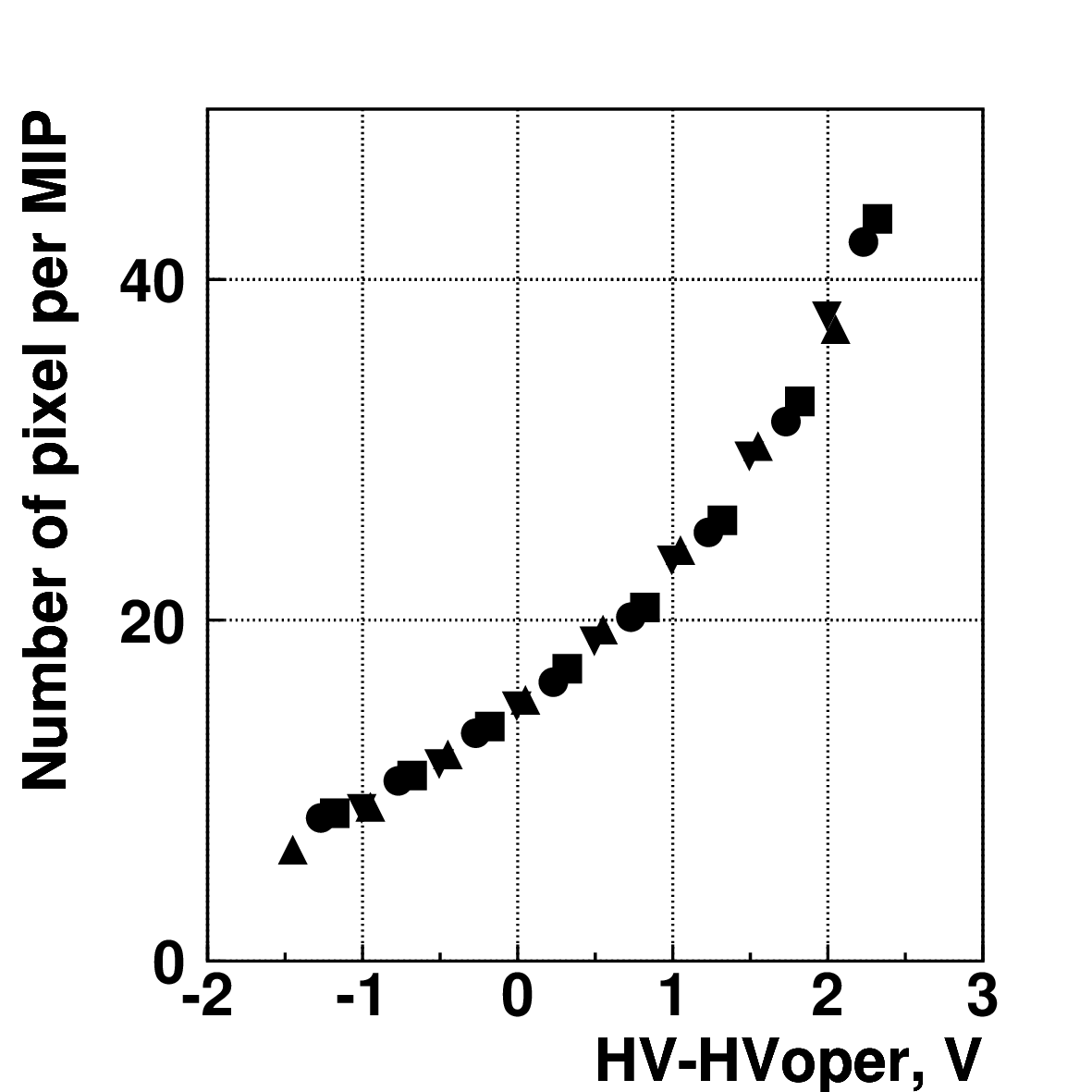}
\includegraphics[width=4.35cm]{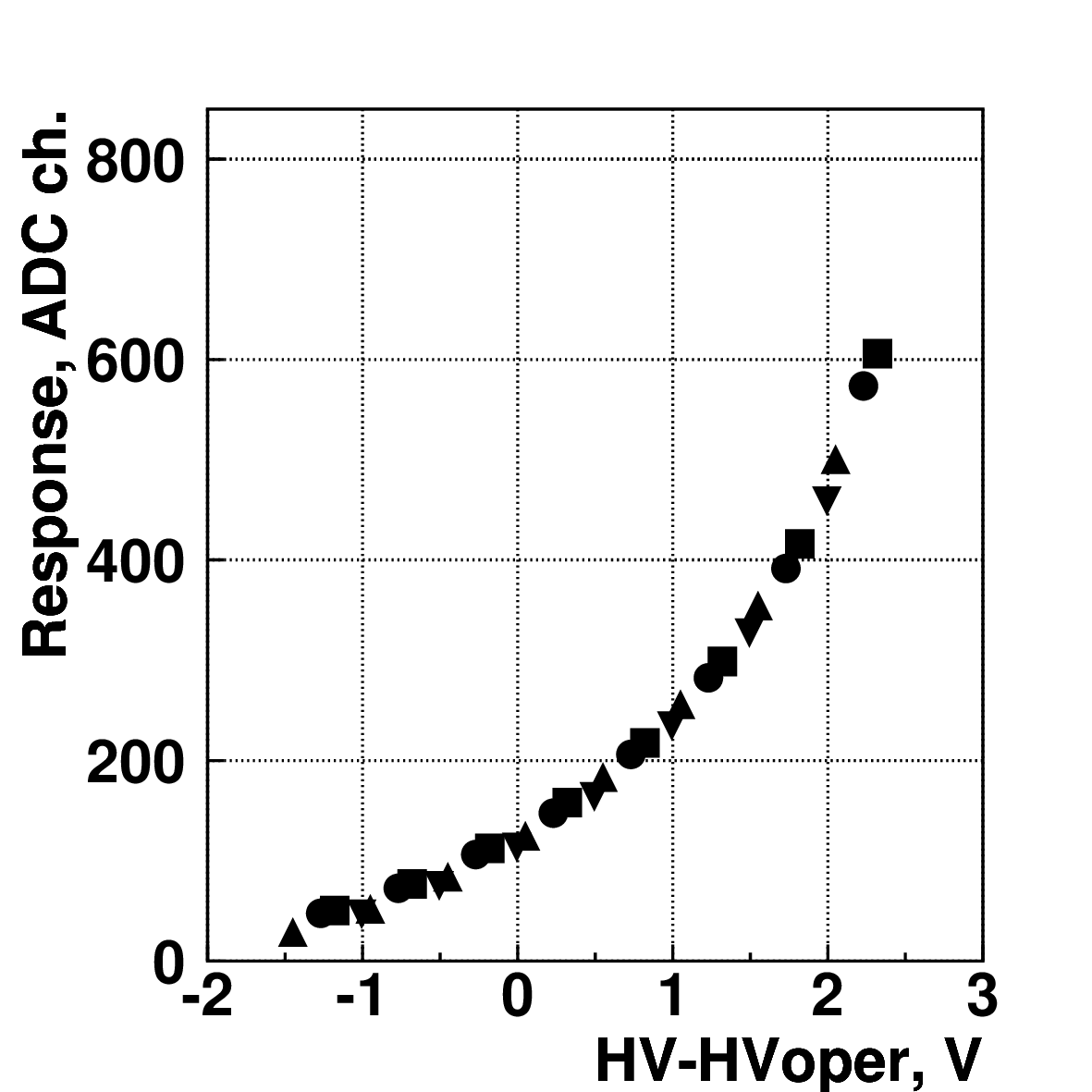}
\includegraphics[width=4.35cm]{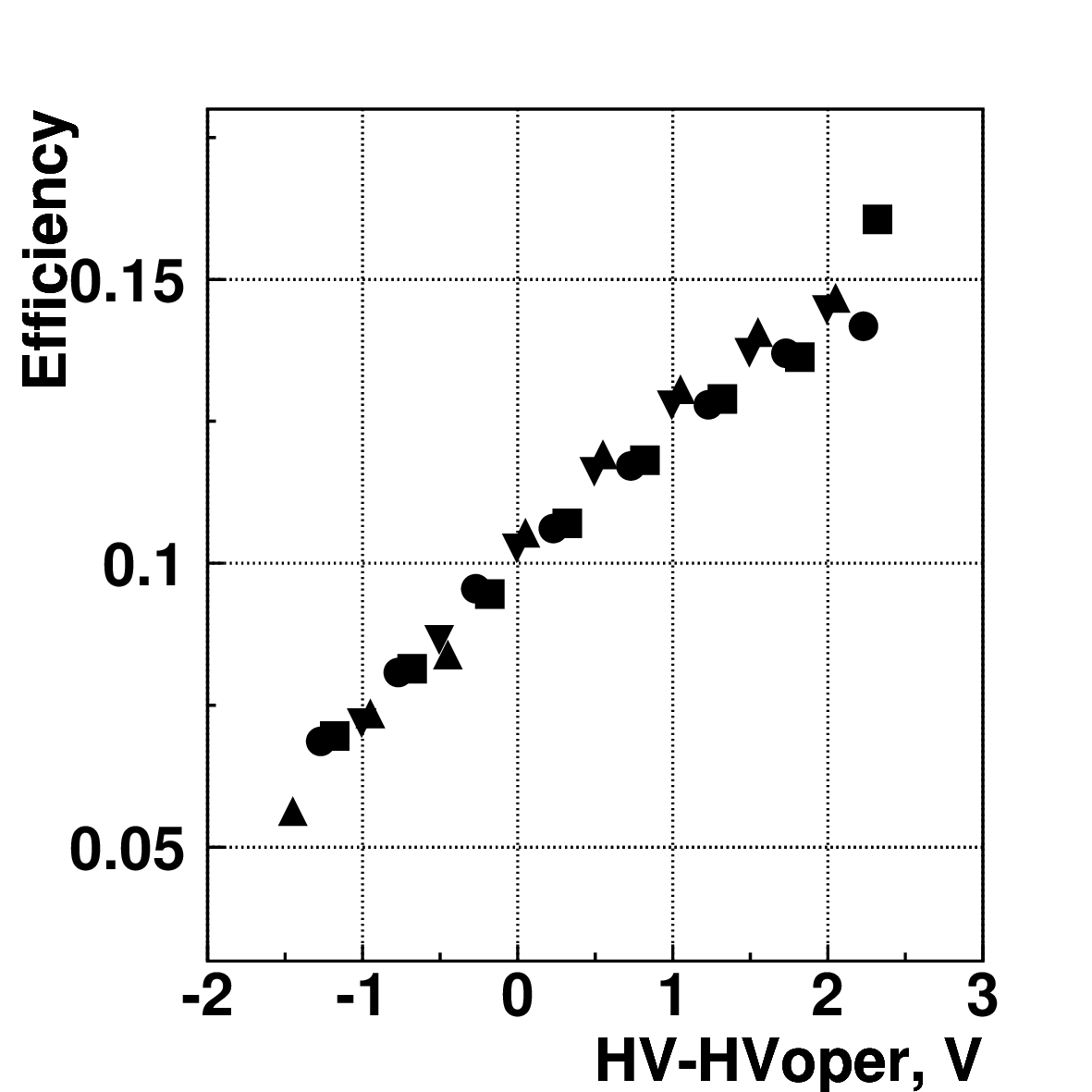}
\includegraphics[width=4.35cm]{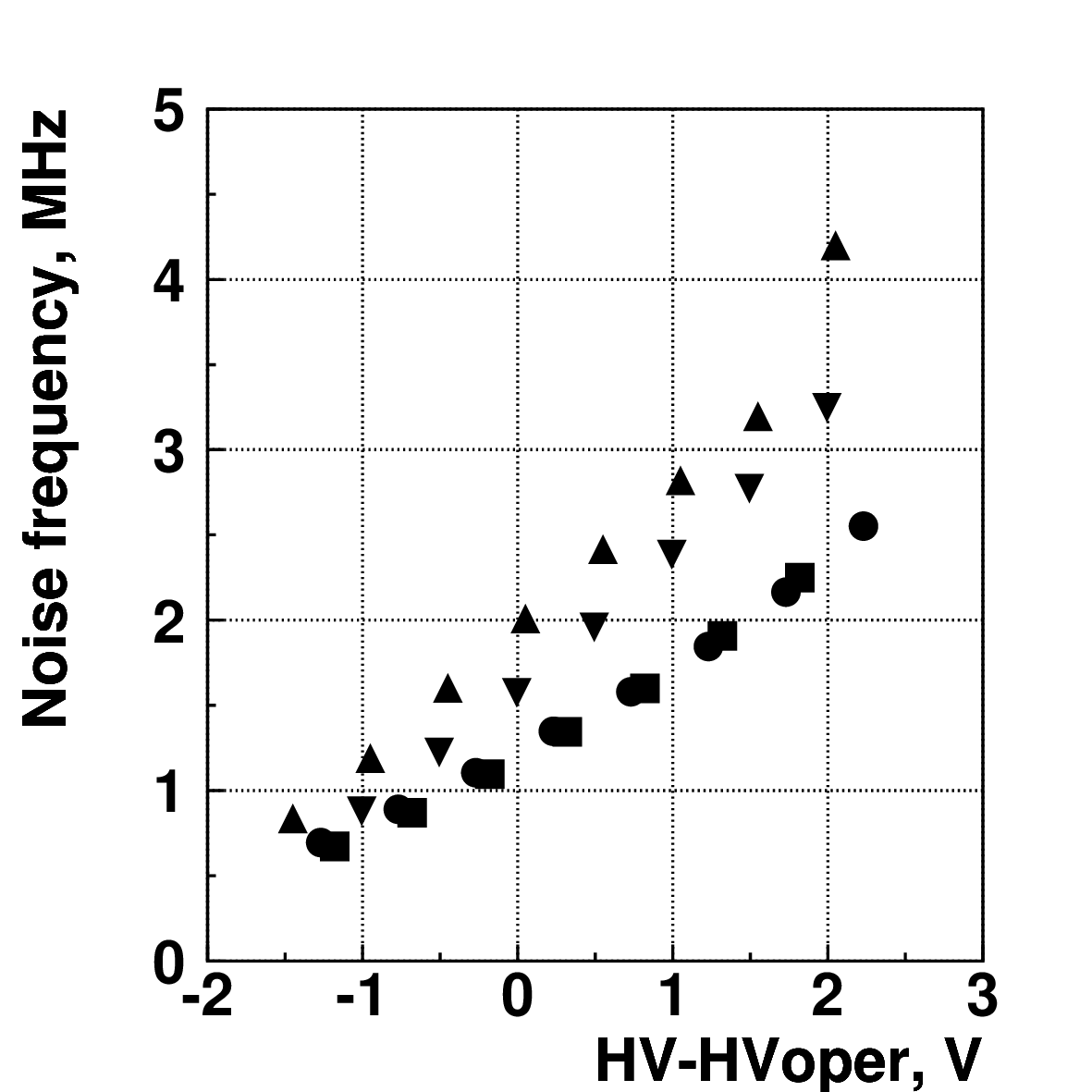}
\caption{MEPhI SiPM parameter dependence on bias
  voltage. Different symbols correspond to 4 randomly selected SiPMs.}
\label{SiPMparam}
\end{figure}

$A=G\times N_{\gamma}\times PDE\times(1+XT) \times(1+AP),$\\ where
$XT$ is the cross-talk probability, and $AP$ is the after-pulse
probability multiplied by the average after-pulse amplitude
suppression.  For large light signals the saturation due to limited
number of pixels becomes important.  Since the $PDE$, $G$, $XT$, and
$AP$ all grow with $\Delta V$, the SiPM response grows very
non-linearly. Fig.~\ref{SiPMparam} shows MEPhI SiPM parameter
dependence on the bias voltage. 
%There is a considerable spread between
%different SiPMs in the gain and cross-talk. The spread of other
%parameters is small.
%
The MEPhI SiPM response at the working point selected for the
CALICE application varies by $6\%/0.1\,$V and $-3.5\%/^\circ
C$\cite{tarkov}. At higher $\Delta V$ the sensitivity of the SiPM
response to variations of the bias voltage and temperature is weaker.
\subsection{Dark rate}
SiPMs have a quite high dark rate originating from charge carriers
thermally created in the sensitive volume. The typical dark rate is
1-2MHz/mm$^2$ at room temperature. The dark rate decreases by a factor
of 2 with the temperature decrease by $8^\circ C$. Hamamatsu SiPMs
have considerably smaller dark rate.
The dark rate grows linearly with the $\Delta V$ because of the
increase in the probability for charge carriers to trigger the Geiger
discharge.
\subsection{Time resolution}
The SiPM response is intrinsically very fast due to a very fast Geiger
discharge development in a thin ($\sim$1-2$\,\mu$m) depletion
layer. The single p.e. timing resolution of about 0.1ns has been
observed\cite{BorisRev}. The timing resolution improves as
$1/\sqrt{N_{p.e.}}$.
\subsection{Insensitivity to magnetic field}
SiPMs are not sensitive to  a magnetic field. This was tested up to 4T
for the MEPhI SiPMs. The SiPM response, gain, cross-talk, and noise
frequency did not change in the magnetic field within the measurement
accuracy\cite{MINICAL}. This feature is extremely important for many
SiPM applications.
\subsection{Radiation hardness}
\begin{figure}[htbp]
\includegraphics[width=5cm]{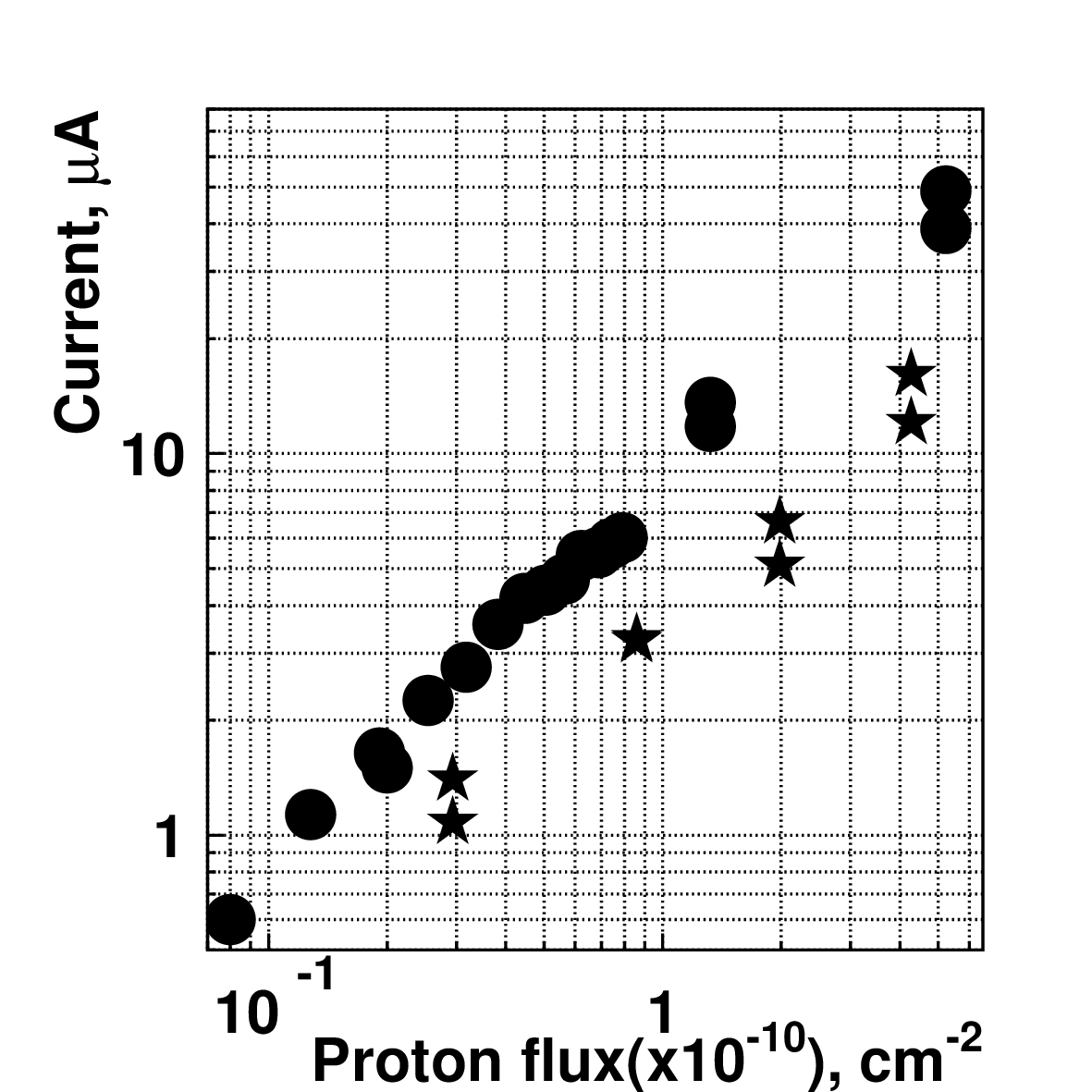}
\caption{The dark current increase with the proton flux for MEPhI
  SiPMs (200$\,$MeV protons, circles) and CPTA MRS-APD (80$\,$MeV
  protons, stars) operated at the same $PDE$. Each star corresponds to
  a different MRS-APD measured after about 1 day after the
  irradiation. Circles below $10^{10}$ flux show the current of one
  SiPM measured online. Other circles correspond to different SiPM
  samples measured about 1 day after the irradiation.}
\label{pirrad}
\end{figure}

The SiPM dark current grows linearly with the particle flux as in
other Silicon detectors\cite{danilov-eigen}. However, since the
initial single photoelectron resolution of SiPMs is by far better than
that of say APD, it starts to suffer earlier. The radiation induced
dark current in a SiPM is described by the following formula:

$I= K\cdot F\cdot D\cdot G\cdot
P_G\cdot(1+XT)\cdot(1+AP)\cdot S\cdot\epsilon \cdot L_{eff},$\\
where F is the particle flux, $K=6\times10^{-17}$A/cm\cite{Lindst}, $S$
is a SiPM area, $L_{eff}$ is
the effective thickness from which charge carriers are collected, $D$
is the energy dependent conversion factor of radiation damage of
different particle species to that of 1$\,$MeV neutrons.
Fig.\ref{pirrad} shows the dark current dependence on proton fluence
for MEPhI and CPTA-149 SiPMs at the same $PDE$ ($\sim10\%$) for
the green light.
The CPTA detectors show about 2 times smaller current increase.
Moreover they were irradiated with 80MeV protons which have $D\sim 2$ while the
MEPhI SiPMs were irradiated with 200MeV protons which have $D\sim 1$.
Therefore the effective thickness of the sensitive layer derived from
this measurements is 5 times smaller for the CPTA detector. It is
about $5\mu$m while the MEPhI SiPM has $L_{eff}\sim25\,\mu$m.  One
can conclude that the CPTA MRS-APD is less vulnerable to the radiation
damage. However, if one operates the CPTA MRS-APD at $PDE$=30\% the dark
current increases even a bit faster than in the MEPhI SiPM.

The annealing effect at a room temperature after the proton
irradiation is small. The current drops by about 35\% in 30 days after
the irradiation with $5\times 10^{10}$ protons/cm$^2$ with
E=80MeV. Relative annealing speed does not depend on the dose up to
this flux which is equivalent to $10^{11}$ $1\,$MeV neutrons/cm$^2$.

When the dark current reaches $\sim5\,\mu$A individual photo-electron
peaks in the SiPM response become smeared. However SiPM can be
operated at much higher currents.

It was demonstrated\cite{MusienkoNDIP} that the gain, $PDE$, $V_{BD}$,
$R$, and a pixel recovery time of SiPMs do not change after the
irradiation with $10^{10}$/cm$^2$ protons with E=82MeV which is
equivalent to $2\times10^{10}$/cm$^2$ of $1\,$MeV neutrons. This was
checked for 5 types of SiPMs from different producers.

\begin{figure}[htbp]
\includegraphics[width=5cm]{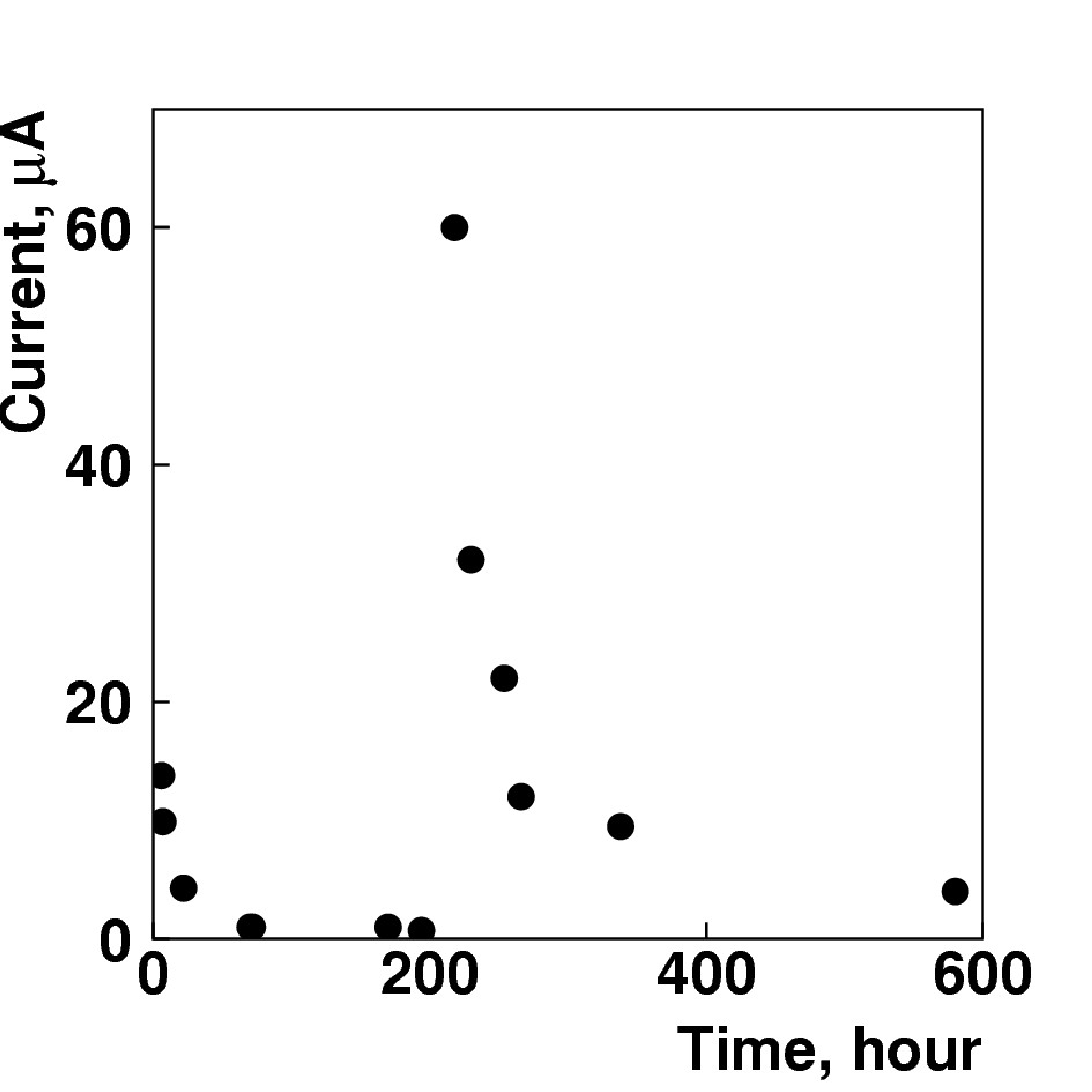}
\caption{The dark current of 16000 pixel MPPC irradiated with a
  $^{60}Co$ source at t=0 (200kRad) and t=200 hours (200 kRad).}
\label{eirrad}
\end{figure}

SiPMs are less sensitive to electromagnetic radiation. For example 10
SiPMs from MEPhI, CPTA, and Hamamatsu were irradiated with a $^{60}Co$
Source\cite{tarkov}. All 10 were still operational after 200krad
irradiation. The dark current of 1600 pixel MPPC(11-025M) increased
considerably after the irradiation but dropped to reasonably small
values after annealing at a room temperature (see
Fig.\ref{eirrad}\cite{tarkov}). The dark current was smaller
(1-3$\mu\,$A after 500krad) for 9 other SiPMs including 400 pixel
MPPC(S10362-11-050U). They were operational even after a $600\,$kRad
dose.  The reason for the fast increase of the current in 1600 pixel
MPPC is not clear.  One should keep in mind that only one sample was
irradiated and more systematic tests are needed. The large dependence
of the radiation hardness on the SiPM type (if confirmed) would
require detailed radiation studies for each SiPM type.
\subsection{Long term stability}
The CALICE Hadronic calorimeter with 7620 MEPhI SiPMs has been
operated at CERN and FNAL during more than 6 months.
%\cite{Erika}
We have not observed any increase in the number of dead SiPMs within
the measurement uncertainty of about 0.1\%. We also have not observed
any major change in the SiPM parameters. However a more
quantitative analysis of their performance is still to be performed.
\subsection{Comparison of SiPMs used in mass applications}
\begin{figure}[htbp]
\includegraphics[width=9cm]{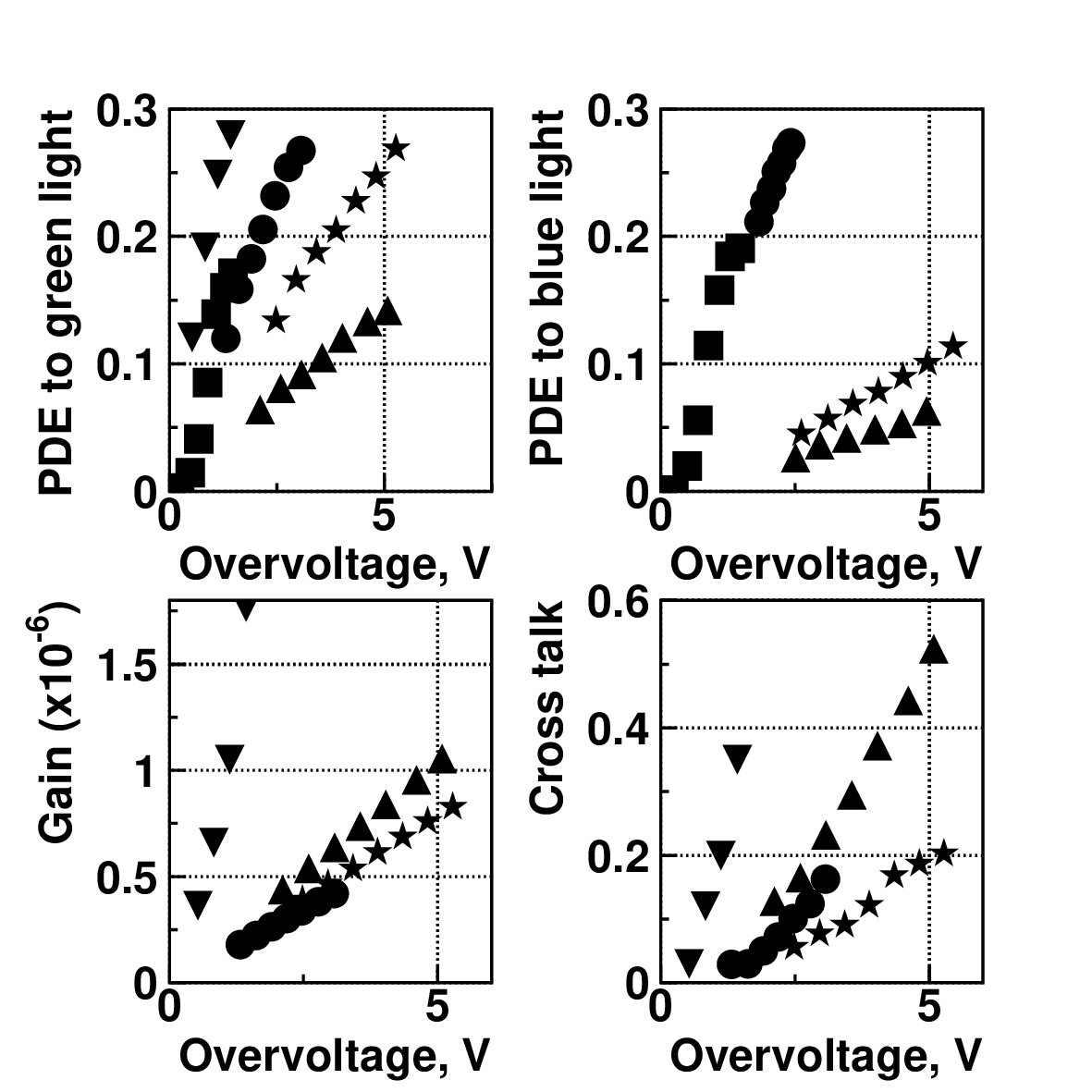}
\caption{ Parameter dependence on over-voltage for various
  SiPMs:  $\bullet$ - 1600 pixel MPPC; $\blacktriangledown$ - 400 pixel MPPC;
 $\blacktriangle$ - MEPhI-Pulsar SiPM; $\star$ 556 pixel MRS APD;
 $\blacksquare$ - $2\times2\,$mm$^2$ blue sensitive MRS APD.}
\label{comparison}
\end{figure}

So far SiPMs from 3 producers have been used in quantity for
experiments: MEPhI SiPMs (CALICE Hadronic calorimeter\cite{AHCAL,Felix}),
Hamamatsu MPPCs (CALICE Electromagnetic calorimeter\cite{Toru}) and
CPTA MRS APDs (ALICE TOF test set-up\cite{Akindinov}). Properties of
these photo-detectors are compared in
Fig.~\ref{comparison}\cite{tarkov}.
$PDEs$ were measured for the plastic scintillator (2.5\% PTP and 0.1\%
POPOP) and the Kuraray Y11 WLS fiber emission spectra. They are
labeled as efficiencies for ``blue'' and ``green'' light.  $PDEs$ of MPPC
and MRS APD are very similar for the green light. They are
considerably higher than the $PDE$ of the MEPhI SiPM. The blue enhanced
MRS APD has the $PDE$ similar to MPPC one but it has much higher dark
noise rate. The cross-talk is comparable in the 1600 pixel MPPC and
MRS APD. It is much higher in the MEPhI SiPM and 400 pixel MPPC. So
the MEPhI SiPM has worse parameters than the more modern MPPC and MRS
APD. Nevertheless it was already adequate for mass applications in
calorimetry.
\section{Examples of SiPM applications}
%
%We discuss SiPM applications in particle physics and partially in
%astro-particle physics. Medical applications are discussed in
%different talks at this Conference.
% In particular SiPM is considered
%to be the best photo-detector for PET\cite{PET}.
%
\subsection{Calorimetry}
A small 108 channel hadron calorimeter prototype for the International
Linear Collider (ILC) was the first “mass” application of SiPMs in a
real experiment\cite{MINICAL}. It consisted of
$5\times5\times0.5\,$cm$^3$ scintillator tiles with WLS fiber and
MEPhI SiPMs installed directly in the tile.
%Calorimeter planes with 9 tiles were produced and tested
%at ITEP and transported to DESY for tests at the 6GeV electron beam. 
The signals of about 20 p.e./MIP were transported directly without any
preamplifier to the LeCroy 2249A ADCs via 25m long coaxial
cables. Since there was only very limited experience with SiPMs at
that time two identical calorimeters with different photo-detectors
were tested simultaneously. One of them used a well established
technique of MAPMTs. Another one was based on a relatively new APD
readout. All three calorimeters demonstrated adequate and practically
identical performance. Results from the calorimeter with the novel
SiPM read-out were obtained first because the calibration was based on
the response measurements using the distance between signals with
different number of p.e. which are well resolved by
SiPMs\cite{MINICAL}.  Results from the calorimeter with MAPMT readout
were obtained also fast\cite{MINICAL}. Analysis and calibration of the
APD data took much longer\cite{AndreevAPD}. The very encouraging
experience with the novel SiPM readout resulted in a selection of
SiPMs as a baseline for the ILC analogue hadron calorimeter.

The CALICE scintillator hadron calorimeter prototype consists of 7620
Scintillator tiles with WLS fibers and MEPhI
SiPMs\cite{AHCAL,Felix}.  
%(see Fig.\ref{tile}
%\begin{figure}[htbp]
%\includegraphics[width=9cm]{tile.eps}
%\caption{The CALICE hadron calorimeter tile with installed SiPM and
% the calorimeter plane with tiles of different sizes.}
%\label{tile}
%\end{figure}
All SiPMs were thoroughly tested before installation into the
tiles\cite{tarkov,Rusinov}. The bias voltage was adjusted to get the
same number of pixels per MIP. SiPMs which had too large cross-talk or noise, too low gain or fluctuating dark current were
rejected. Many other parameters were also measured but they were found
to be mainly within the required limits and practically did not result
in additional rejection.

The calorimeter was tested during 2006-2008 at CERN and FNAL.
%\cite{Erica}
It worked very reliably practically without problem during more than 6
months. Only about 1\% of SiPMs were not operational. Majority of them
were from the initial production when the selection procedure was not
yet fully developed. We can conclude that the experience with the
first mass use of SiPMs in the real experiment is very encouraging.

The next engineering prototype is being built now\cite{Felix}. It will
have readout chips installed on very compact PCBs inside the active
calorimeter layer. The scintillator tiles with WLS fibers will be only
3mm thick in order to minimize the gaps in the absorber. We plan to
use CPTA MRS APDs. They have a higher $PDE$ and smaller cross-talk than
the MEPhI SiPMs used in the present prototype (see
Fig.\ref{comparison}). The engineering prototype will be scalable for
the mass production of a few million channel CALICE hadron
calorimeter.

The CALICE collaboration investigates also a possibility to use blue
sensitive SiPMs for a direct read-out of scintillator tiles without
WLS fibers.

The CALICE scintillator-tungsten electromagnetic calorimeter prototype
is based on $1\times4.5\times0.3\,$cm$^3$ scintillator strips with WLS
fiber and the Hamamatsu 1600 pixel MPPC\cite{Toru}.
%type?
The first prototype with $\sim500$ channels was successfully tested at
a 6GeV electron beam at DESY. The second prototype with $\sim2000$
channels is tested at FNAL. Initial experience is again very
encouraging.
\subsection{Muon systems}
Scintillator strips with WLS fiber and MAPMT readout is a well
established technique for muon detection. This technique was
successfully used in the MINOS and OPERA neutrino detectors. It was
shown that by switching to a SiPM readout one gets more p.e./MIP and
simplifies the technique
considerably\cite{balagura,LCWS04MVD}. Because of insensitivity to a
magnetic field this technique can be used in collider detectors. A new
muon and $K_L$ end cap detector is designed now for the SuperBelle
experiment. It will consist of 28 thousand scintillator strips up to 3
meter long\cite{Pakhlov}. We plan to use the CPTA MRS-APDs as the
photo-detectors. It will provide more than 10p.e. from the far end of
$300\times2.5\times1\,$cm$^3$ scintillator strips and a small noise
rate.

The time resolution of about 1ns allows the determination of the
coordinate along the strip with about 15cm
accuracy. Similar approach can be used for the muon
system of the future International Linear Collider\cite{LCWS04MVD}.

Using two SiPMs per a scintillator tile it is possible to build a muon
system with a negligible noise rate. This feature is very useful for
cosmic muon test set-ups. The cosmic ray test set-up for the ALICE TOF
system is based on $15\times15\times1\,$cm$^3$ scintillator tiles with
the WLS fiber and two MRS APD read out\cite{Akindinov}. About 500 MRS
APDs (including spare modules) are used in this effectively working
system.
     
\subsection{Neutrino and Astro-particle applications}

The T2K neutrino oscillation experiment in Japan plans to use SiPMs
practically in all subsystems\cite{T2K}. Altogether about 50 thousand
Hamamatsu MPPCs (S10362-13-050C) will be used. More than 30 thousand
of them have been already produced and about 15 thousand have been
tested. Tested MPPCs show very good uniformity of parameters. The experiment plans to start data taking already in
2009.

SiPM applications in astro-particle physics are discussed in other
talks at this conference. I will mention only two
examples.  The PEBS balloon experiment
%\cite{PEBS}
plans to check an indication of the cosmic positron flux excess due to
Dark Matter annihilation. Linear SiPM arrays will be used in the PEBS
scintillating fiber tracker\cite{PEBS}. Each tracker plane consists of
5 layers of $250\,\mu$m diameter scintillating fibers. The tracker will
have 55 thousand read-out channels. The 32 channel SiPM linear arrays 
from Hamamatsu and IRST are tested. 
% (see Fig.\ref{linarray})
About 10 p.e. per MIP have been observed with the Hamamatsu
device. This resulted in a 89$\mu$m spatial resolution.

The tungsten-scintillator electromagnetic calorimeter will have about 2
thousand $840\times8\times2\,$mm$^3$ scintillator strips with WLS fibers. We plan to use the CPTA MRS APDs for the read-out. 
We have observed about 10 p.e./MIP in the 2mm thick strips.
      
The excellent single photon resolution, high quantum efficiency, low
mass, and low bias voltage make SiPMs an interesting alternative to
standard PMTs in Dark Matter detection in liquid xenon. The interest
to this approach increased after the observation of unexpectedly high
SiPM efficiency of $\sim 5.5\%$ to Xe UV scintillation
light\cite{Aprile}. Unfortunately our measurements give more than an
order of magnitude lower efficiency\cite{Akimov}. Nevertheless we
continue this R\&D but with the wavelength shifters which transform
Xe scintillation light into the SiPM sensitivity region.

\subsection{Time of Flight and Cherenkov counters}

The excellent single p.e. time resolution allows use of SiPMs for
Time of Flight measurements. The timing resolution of 32ps has been
obtained using a $3\times3\,$mm$^2$ MEPhI SiPM coupled to a
$3\times3\times40\,$mm$^3$ BC143 plastic
scintillator\cite{BorisRev}. This resolution contains the contribution
from the scintillator 1.4ns decay time.

At the first glance SiPMs are not suitable for the detection of
individual photons in Cherenkov light rings because of the high noise
rate. However beautiful Cherenkov light rings have been observed
recently\cite{Krizan}. The number of p.e. per ring was larger than
with MAPMTs. The SiPM noise was reduced using the excellent SiPM
timing resolution.
    
\subsection{Medical and other applications}

Medical applications of SiPMs are discussed at this Conference by
A. Del Guerra\cite{PET}.  We would like to mention only the
conclusion of this talk - SiPMs are the most promising photo-detectors
for the Positron Emission Tomography. It is possible to anticipate
other applications of SiPMs in medicine.

The compactness, low bias voltage, large output signals due to the
Geiger amplification, single photon counting capability, high quantum
efficiency, insensitivity to magnetic field, excellent time
resolution, and low cost make SiPM attractive for many other
application, for example for radiation monitoring, compact dosimeters
etc.
 
\section{Conclusions}

SiPMs have many advantages over usual PMTs and other Si
detectors. Their basic properties are relatively well
understood. SiPMs are already produced by many companies. More than 10
thousand SiPMs have already been used in real experiments and
demonstrated excellent performance. Several experiments in particle
physics plan to use tens of thousand SiPMs each. So in the near
future the number of SiPMs used in experiments will be comparable to
the number of the used APDs.  We anticipate a wide use of SiPMs in
other fields in particular in medicine. There is a very fast
development of new and better SiPMs. In spite of several limitations
like a small sensitive area and a large noise, SiPMs will become one
of the most popular photo-detectors.

 \section{Acknowledgements}

We are grateful to many people for useful discussions in particular to
A.Akindinov, V.Balagura, B.Dolgoshein, E.Garutti, V.Golovin, S.Klemin,
R.Mizuk, Yu.Musienko, P.Pakhlov, E.Popova, V.Rusinov, F.Sefkow, E.Tarkovsky,
I.Tikhomirov. It was a pleasure to work together with many co-authors
of the studies reported in this paper. This work was supported in part
by the grants RFBR1329.2008.2, RFBR08-02-12100, RFBR/Helmholtz
HRJRG-002 and SC Rosatom.
% \appendix
% \section{}
% \label{}

\end{document}